# On the vacancy decay in endohedral atoms

M. Ya. Amusia[1,2] and A. S. Baltenkov[3]

[1]Racah Institute of Physics, The Hebrew University, Jerusalem 91904, Israel
[2]Ioffe Physical-Technical Institute, St.-Petersburg 194021, Russia
[3]Arifov Institute of Electronics, Tashkent, 700125, Uzbekistan

**Abstract**

It is demonstrated that the fullerene shell affects dramatically the radiative and Auger vacancy decay of an endohedral atom $A@C_{60}$. The collectivized electrons of the $C_{60}$ shell add new possibilities for radiative and non-radiative decays similar to that in ordinary atoms where initial and final state vacancies almost always belong to different subshells.

It is shown that the smallness of the atomic shell radii as compared to that of the fullerenes shell provides an opportunity to derive the simple formulas for the probabilities of the electron transitions. It is shown that the radiative and Auger (or Koster-Kronig) widths of a vacancy decay due to electron transition in the atom A in $A@C_{60}$ acquire an additional factors that can be expressed via the polarizability of the $C_{60}$ at transition energy.

It is demonstrated that due to opening of the non-radiative decay channel for vacancies in subvalent subshells the decay probability increases by five – six orders of magnitude.

## 1. Introduction

It is already more than thirty years since it was predicted that intra- and inter-shell interactions in atoms could modify considerably the probability of the radiative and Auger-decay of vacancies in atoms [1]. It was demonstrated that due to this interaction the decay probability could increase or decrease, up to complete disappearance of the decay photon or electron line. A new phenomenon was predicted, namely radiative self-locking of atomic shells, the essence of which is complete suppression of the radiative decay of some atomic vacancies due to interelectron interaction [1, 2].

Since then, new multi-atomic and hence multi-electronic structures, such as metallic clusters and fullerenes, became available and started to be objects of intensive investigations. There electron correlation effects in all processes, including vacancies decays, could be much stronger than in isolated multi-electron atoms.

Recently in a number of papers (see e.g. [3-5] and references therein) it was demonstrated that in endohedral systems $A@C_{60}$, in which atom A is located inside the "empty" inner part of the $C_{60}$, the photoionization characteristics of atom A are impressively modified by the $C_{60}$ shell.

The aim of this paper is to show that the fullerenes $C_{60}$ electron shell drastically modifies also the decay probability of a vacancy in the atom A in $A@C_{60}$. This modification is a result of a number of effects, including opening of new decay channels of a vacancy and profound modification of those channels that exist in isolated atom A. Namely, the decay can proceed due to opening of new channels, which account for participation of $C_{60}$ electrons. This is similar to the situation in ordinary molecules [6] and clusters [7], where a vacancy in one atom can decay due to transition in another.

It is natural to distinguish decays of vacancies in endohedral atoms by the degree of participation of the $C_{60}$ electrons. Namely, the vacancy in the endohedral atom can be closed by an outer electron either from the same atom, or from the collectivized electrons of the $C_{60}$, forming what could be called *atomic* and *fullerenes* decay, respectively. If atomic vacancy is deep enough, it can be closed be one of the non-collectivized inner electrons of carbon atoms that form the $C_{60}$, presenting *molecular* decay.

If decay results with emission of a photon, the $C_{60}$ shell can intercept it, and then emit a photon with the same energy. As a result, the radiative decay and the decay process becomes interference. Therefore, the decay probability similar to what was predicted in isolated atoms [1] can be altered dramatically, from strong enhancement till almost complete suppression. The same situation can happen in the Auger decay: the interaction between electrons, inner and outer, can be modified due to virtual excitations of $C_{60}$ electrons, thus substituting the pure Coulomb inter-electron interaction $V$ by a transferred energy $\omega$ dependent effective one $\Gamma(\omega)$ (see e.g. [8]). The difference between $V$ and $\Gamma(\omega)$ is determined by so-called intra- or intershell effects.

The physical nature of these so-called intershell effects in the vacancy decay processes has the following meaning. A many-electron atomic subshell is polarized by emission of the decay photon, or in other words, virtually excited by it. Therefore a time-dependent dipole moment is induced. Under the action of this dipole moment a neighboring atomic subshell is ionized. The inclusion of this effect in isolated atoms was performed in [1] in the frame of so-called RPAE [9], which is extremely convenient to describe this effect. Since the electronic subshells in an atom are not spatially separated well enough, the amplitude of such a two-step photo-process cannot be expressed accurately enough directly via the dipole polarizabilities of the many-electron subshells.

The direct application of RPAE to A@$C_{60}$ is an extremely complex task. In fact this is a hard task even for isolated atoms, as it was demonstrated already in Ref. [1]. However in A@$C_{60}$ we have an important simplifying factor. Indeed, the radius of $C_{60}$ shell significantly exceeds that of an encapsulated atom. This makes it possible for vacancy decay of the A atom, in the first approximation, to consider the electronic sub-systems of the fullerene shell and atom as practically independent of each other. For this reason, the amplitude of atomic vacancy decay that goes through virtual excitation of the $C_{60}$ shell electrons can be expressed directly via the dynamic polarizability of the fullerene shell $\alpha_{C_{60}}(\omega)$. In those cases when the decay energy is close to or particularly lover than the frequency of plasma oscillations of the collectivized electrons of the $C_{60}$ shell, the role of this two-step decay becomes decisively important.

The big difference between atomic and fullerenes radii, as shown below, permits deriving the formulas that are valid even beyond the RPAE frame, presenting expressions not for all but in principal for experimentally distinguishable decay channels of vacancies in the A atoms encapsulated in A@$C_{60}$ via experimentally measurable dipole polarizability $\alpha_{C_{60}}(\omega)$ and radiative decay probability of isolated atom A. The present paper is devoted to deriving the formulas connecting the probabilities of radiative and non-radiative decays of the vacancies in the endohedral and free atoms. The method developed can be applied to other objects where many other atoms with collectivized electrons surround a given atom or an ion.

## 2. Essential formulae – radiative decay

The radiative decay amplitude can be presented by a the following diagrammatical relation [1, 9]

$$D_{if}(\omega_{if}) = d_{if} + \chi_{C_{60}}(\omega_{if}) D_{C_{60}}(\omega_{if}) \omega_{if} \quad (1)$$

Here the line with an arrow to the left represents a vacancy in the atom A either in the initial $i$ or in the final state $f$. The dashed line represents a photon; dark circle stands for the vacancy decay $i \rightarrow f$ amplitude $D_{if}(\omega_{if})$ and $\omega_{if}$ is the energy of the emitted photon. The vertically oriented wavy line denotes the potential of the Coulomb interaction between an electron in atom A and in $C_{60}$. The loop $\chi_{C_{60}}(\omega_{if})$ represents the electron-vacancy real or virtual excitation of $C_{60}$. The gray circle is the amplitude $D_{C_{60}}(\omega_{if})$ of the photon $\omega_{if}$ emission. The small dark dot represents the intra-atomic corrections to the amplitude of real or virtual photon emission.

The interaction between atomic and $C_{60}$ electrons is included in the lowest order, since the radius R of $C_{60}$ is much bigger than the radius $\rho$ of the atom A and the thickness $t_{C_{60}}$ of the fullerene $C_{60}$. So, the addition of each extra interaction line leads to a small factor $\rho/R \ll 1$ in the corresponding contribution to the amplitude and as such can be neglected. Note that due to the same reason and with the desire to obtain qualitatively sound results we neglect exchange between atomic and $C_{60}$ electrons.

Representing the Coulomb interaction $V = |\mathbf{r}_a - \mathbf{r}_{C60}|^{-1}$* between an atomic electron with coordinate $\mathbf{r}_a$ and a fullerene shell electron with coordinate $\mathbf{r}_{C60}$ by the first term that leads to non-vanishing contribution to $D_{if}(\omega_{if})$, one substitutes $V$ by

$$V \approx \mathbf{r}_a \cdot \mathbf{r}_{C60} / R^3. \quad (2)$$

It is taken into account in (2) that $r_a \ll r_{C60} \approx R$. The relation $r_{C60} \approx R$ holds since the integration over $|\mathbf{r}_{C60}|$ takes place within the thickness $t \geq \Delta r_{C60}$ of the $C_{60}$ shell with $t \ll R$.

Then one has for $D_{if}(\omega_{if})$ in the operator form

$$\hat{D}(\omega_{if}) = \hat{d} + \hat{d}\hat{\chi}_{C_{60}}(\omega_{if})\hat{D}(\omega_{if})/R^3. \quad (3)$$

---

* The atomic system of units: $e = m = \hbar = 1$ is used throughout this paper

Here the "loop" $\hat{\chi}_{C_{60}}(\omega_{if})$ or the propagator of $C_{60}$ electron excitation, i.e. electron-vacancy pair creation is given by the relation $\hat{\chi}_{C_{60}} = 1/(\omega - \hat{H}_{ev}) - 1/(\omega + \hat{H}_{ev})$ [9], and $\hat{H}_{ev}$ is the electron-vacancy pair Hartree-Fock Hamiltonian of $C_{60}$.

Then one has to use the definition of the dipole polarizability $\alpha(\omega)$ of a multi-electron system (see e.g. [8, 9])

$$\alpha(\omega) \equiv -[2 \sum_{evexit} \omega_{ev} D_{ev}(\omega)(\omega^2 - \omega_{ev}^2)^{-1} d_{ev}]. \tag{4}$$

In (4) we employ an alternative definition of $\alpha(\omega)$. Indeed, usually $\alpha(\omega)$ is defined as

$$\alpha(\omega) \equiv -[2 \sum_{evexit,o} \omega_{ev} |D_{ev}(\omega_{ev})|^2 (\omega^2 - \omega_{ev}^2)^{-1}], \tag{5}$$

but it can be easily demonstrated that this definition and that in (4) are identical [9]. Then we obtain the following relation for the radiative decay amplitude

$$D_{if}(\omega_{if}) = d_{if}\left[1 - \frac{\alpha_{C_{60}}(\omega_{if})}{R^3}\right], \tag{6}$$

which leads to a remarkably simple formula for the radiative width

$$\Gamma_{\gamma,if}^{C60} = \Gamma_{\gamma,if}^{A} \left|1 - \frac{\alpha_{C_{60}}(\omega_{if})}{R^3}\right|^2. \tag{7}$$

Note that a factor at $\Gamma_{\gamma,if}^{A}$, $G_{C_{60}}(\omega) = \left|1 - \alpha_{C_{60}}(\omega)/R^3\right|^2$, was recently introduced in [8] while discussing the role of $C_{60}$ electron shell upon photoionization of the endohedral atom A@$C_{60}$. Similar factor in pure classical derivation was obtained in Ref. [10].

## 3. Auger decay

In this section we will start with discussing Auger-decay that became possible in A@$C_{60}$ while is energetically forbidden in isolated atoms A. We will concentrate only on dipole transitions $i \to f$, because only for them the radiative decay channel in A is open.

The diagram that describes this process is depicted in the following way

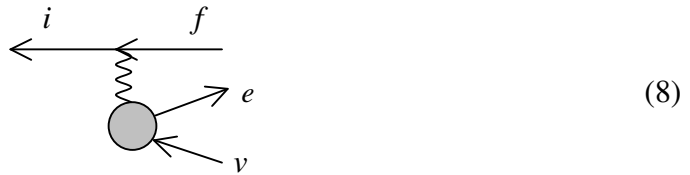 (8)

Here $ev$ is an electron-vacancy pair that belongs to $C_{60}$. As in the previous section, we can consider the process in the lowest order in interaction between atomic and $C_{60}$ electrons, neglect exchange and use (2) for the inter-electron interaction represented by the wavy line. For the Auger-decay is more convenient to receive an analytical expression for the Auger decay probability that is given by the square modulus of (7) summed over all possible final states. Then one has for the Auger - decay width $\Gamma_{A,if}^{C60}$ with atomic vacancy $i$ transition into $f$ in A@$C_{60}$:

$$\Gamma_{A,if}^{C60} = r_{if}^2 \frac{1}{R^6} \operatorname{Im} \alpha_{C_{60}}(\omega_{if}). \qquad (9)$$

Here $r_{if}$ is the radial matrix element between states of the atom A – $i$ and $f$. The expression (9) can be conveniently transformed into a relation that connects only directly measurable characteristics of the isolated atom A and $C_{60}$:

$$\Gamma_{A,if}^{C60} = \Gamma_{\gamma,if}^{A} \frac{3}{2}\left(\frac{c}{\omega_{if}}\right)^3 \frac{\operatorname{Im}\alpha_{C_{60}}(\omega_{if})}{R^6} = \Gamma_{\gamma,if}^{A} \frac{3}{8\pi}\left(\frac{c}{\omega_{if}}\right)^4 \frac{\sigma_{C_{60}}(\omega_{if})}{R^6}, \qquad (10)$$

where $\sigma_{C_{60}}(\omega_{if})$ is the total photoabsorption cross-section of $C_{60}$. If one is interested to have the Auger width that correspond to one-electron ionization channel, $\sigma_{C_{60}}(\omega_{if})$ has to be substituted in (10) by $\sigma_{C_{60}}^+(\omega_{if})$ - the one-electron photoionization cross-section that can be found in Refs. [11-13].

Note, that if $i$ and $f$ vacancies belong to different subshells of the same shell, the transition presented by (8) is called Koester – Kronig transition. Since $\operatorname{Im}\alpha_{C_{60}}(\omega)$ rapidly decreases with $\omega$ growth, $\Gamma_{A,if}^{C60}$ is considerable for $\omega \leq 50 eV$ that is for Koester – Kronig transitions only.

An important characteristic of vacancy decay is the so-called fluorescence yield $J_R$, which is the ratio of the radiative $\Gamma_{rad}$ and the total $\Gamma_{tot} = \Gamma_A + \Gamma_\gamma$ widths. Dividing (7) only by (10), since in our case $\Gamma_{\gamma,if}^{C60} \ll \Gamma_{A,if}^{C60}$, one has

$$J_R(\omega_{if}) = \frac{2}{3}\left(\frac{\omega_{if}}{c}\right)^3 \frac{R^6}{\operatorname{Im}\alpha_{C_{60}}(\omega_{if})}\left|1 - \frac{\alpha_{C_{60}}(\omega_{if})}{R^3}\right|^2 = \frac{8\pi}{3}\left(\frac{\omega_{if}}{c}\right)^4 \frac{R^6}{\sigma_{C_{60}}(\omega_{if})}\left|1 - \frac{\alpha_{C_{60}}(\omega_{if})}{R^3}\right|^2. \qquad (11)$$

If $i \to f$ is not a dipole, but a monopole $l = 0$ or quadrupole $l = 2$ transition, the Auger- width is given by expression similar to (9), $\Gamma_{Aif(l)}^{C60} \sim |(r^2)_{if}|^2 \operatorname{Im}\alpha_{C_{60}}^l(\omega)/R^{10}$ that is by the factor $\rho^2/R^2 \ll 1$, where $\rho$ is the atomic radius, smaller than the dipole decay width given by (9). Since nothing is known about non-dipole polarizabilities of the fulleren $C_{60}$, we cannot present numerical results for the widths of these transitions.

Let us turn now to the Auger-decay of those vacancies, for which it is allowed also in the isolated atom A. In this case the process has two pathways – direct decay with emission of atomic outer electron and indirect that leads to the same atomic electron emission via virtual or real excitation of the $C_{60}$ electrons

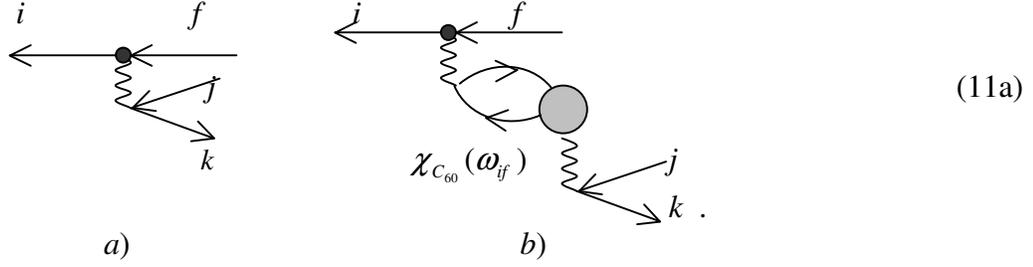

(11a)

a)   b)

Here $k$ and $j$ are the electron-vacancy excitations of the atom A that are formed in Auger-decay. The dark dot includes all virtual intra-atomic virtual excitations.

In this paper we concentrate only on dipole transitions $i \to f$, since we know well the dipole excitation spectrum from experiment, including its main feature from the vacancy decay probability point of view, namely the Giant resonance. The two diagrams of (11a) present two pathways that lead to decay. They can interfere leading both to enhancement and suppression of the decay probability. Obviously, the contribution of (11b) is particularly important when $\omega_{if}$ is close to the Giant resonance frequency. It means that important $C_{60}$ influence can be expected, as it was mentioned above while discussing equation (10), for Koester-Kronig (super – Koester – Kronig) transitions, in which vacancy $f$ ($f, j$) belong to the same shell as $i$.

Analytically, the amplitude $A_{i \to fkj}$ of (11a) can be presented using (2) in the following form:

$$A_{i \to fkj} = A^A_{i \to fk} - r^A_{if} \frac{\alpha_{C_{60}}(\omega_{if})}{R^6} r^A_{jk}, \qquad (12)$$

where the upper index A denotes the respective values that belong to the isolated atom A. The relation (12) leads to the following equation for the Auger-width of A@$C_{60}$:

$$\Gamma^{C_{60}}_{A, i \to fkj} = \Gamma^A_{A, i \to fkj} \left| 1 - \frac{1}{2\pi} \sqrt{\frac{3}{2}} \frac{c^2}{\omega_{if}} \frac{\alpha_{C_{60}}(\omega_{if})}{R^6} \left( \frac{\Gamma^A_{\gamma, if}}{\Gamma^A_{A, i \to fkj}} \right)^{1/2} [\sigma^A_j(\varepsilon_k)]^{1/2} \right|^2. \qquad (13)$$

Here $\sigma^A_j(\varepsilon_k)$ is the photoionization cross section of atomic subshell $j$ with emission of a photoelectron with the energy $\varepsilon_k$. Note that the factor in brackets is by the order of magnitude close to one. This is seen from (12), where the first expression can be estimated as $A^A_{i \to fk} \sim r^A_{if} r^A_{jk} / \rho^3$ so that the second term in (12) gives a correction of the order $\rho^3 \alpha_{C_{60}}(\omega_{if}) / R^6 \approx \rho^3 / R^3 \ll 1$.

### 4. Determination of polarizability

Since we do not know the measured values of $\alpha_{C_{60}}(\omega)$, we will find the dipole polarizability of $C_{60}$ via its experimentally known [11-13] photoabsorption cross section $\sigma_{C_{60}}(\omega)$. This method was developed in Ref. [14]. The $\omega$ dependence of photoabsorption cross-section exhibits a powerful maximum at the so-called Giant resonance frequency that is of about 20 eV. This energy is close to the decay energy of a number of radiative and non-radiative atomic transitions.

Using the relation between the imaginary part of the polarizability and the photoabsorption cross-section $\operatorname{Im}\alpha_{C_{60}}(\omega) = c\sigma_{C_{60}}(\omega)/4\pi\omega$, one can calculate the dynamical polarizability of the $C_{60}$ shell. Although the experiments [11-13] provide no direct absolute values of $\sigma_{C_{60}}(\omega)$, it can be reliably estimated by using the different normalization procedures based on the sum rule $(c/2\pi^2)\int_{I_o}^{\infty}\sigma_{C_{60}}(\omega)d\omega = N$, where $N$ is the number of collectivized electrons. The dispersion relation

$$\operatorname{Re}\alpha_{C_{60}}(\omega) = \frac{c}{2\pi^2}\int_{I_{60}}^{\infty}\frac{\sigma_{C_{60}}(\omega')d\omega'}{\omega'^2 - \omega^2} \qquad (14)$$

connects the real part of the polarizability $\operatorname{Re}\alpha_{C_{60}}(\omega)$ with the imaginary part $\operatorname{Im}\alpha_{C_{60}}(\omega)$. In Eq. (14) $I_{60}$ is the $C_{60}$ ionization potential.

This approach was used to derive the polarizability of $C_{60}$ in [8, 14], where it was considered that $N = 240$, i.e. 4 electrons collectivized from each C atom. Using the photoabsorption data considered in [12] as most reliable, we obtained $N_{eff} \approx 250$. This is sufficiently close to the value accepted in [8, 12]. Note that since the one-electron photoionization cross-section of $C_{60}^+$ is much smaller than the similar cross section for $C_{60}$, one cannot limit with this cross-section measured and calculated in [12]. Obviously, other photoionization channels are of importance and have to be included.

The calculation results of the real and imaginary parts of dynamic polarizability of the fullerene shell $C_{60}$ are given in Fig.1. The experimental photoabsorption cross-section for $C_{60}$ taken from [12] is presented in the inset of this figure. As seen from this figure, the cross section is small at threshold (which also means relatively low intensity of discrete excitations) and is dominated by a huge maximum that is called Giant resonance well above the threshold. Therefore, in (14) a small contribution of discrete excitations of the $C_{60}$ collectivized electrons are neglected.

The frequency dependence of the imaginary part, as it should be, is similar to the frequency dependence of $\sigma_{C_{60}}(\omega)$. A small peak in the cross-section for photon energy ~5 eV is transformed into a significant maximum, which is explained by a small value of photon energy as compared to energy of the giant resonance ~22 eV. The real part of the polarizability obtained as a result of integration according to formula (14) behaves more systematically and with the raise of radiation frequency $\omega$, as it should be, decreases as $\operatorname{Re}\alpha_{C_{60}}(\omega) \sim -N_{eff}/\omega^2$. The calculation results of dynamical polarizability will be used further to calculate the ratio of widths of the radiative and non-radiative decays of vacancies in the endohedral atoms.

## 5. Role of Auger-electron reflection

Until now we considered the effect of the fullerene shell on the probabilities of the radiative and non-radiative transitions in the endohedral atoms. However, the $C_{60}$ shell can affect the outgoing Auger-electrons from the A atom in the same way as it affects photoelectrons [4, 5, 15]. This can be taken into account similarly to how it was done in these papers. In the Auger-decay of a vacancy inside the atom A of $A@C_{60}$ strong modification of its probability can come from reflection and refraction of the Auger-electron.

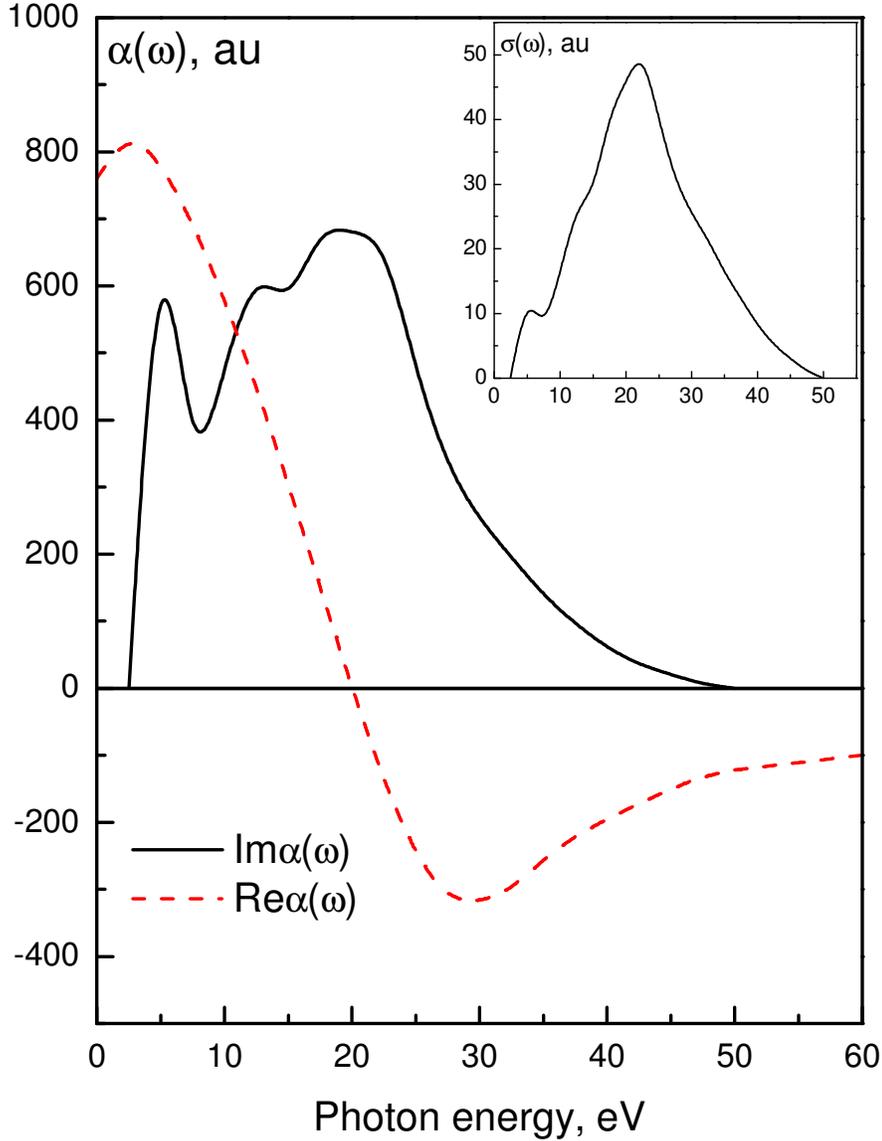

Fig. 1. Real and imagine parts of the dynamical dipole polarizability of $C_{60}$.

In order to take into account these processes for the Auger- electron's wave, we use here, just as in [4, 5, 15] a $\delta$ - bubble model that represents the static $C_{60}$ potential as $U(r) = -V_0 \delta(r - R_c)$, with $V_0$ chosen in such a way as to reproduce the binding energy of $C_{60}^-$ negative ion. In the frame of this model potential, its influence upon the photoionization amplitude is presented by a factor $F_A(\varepsilon)$, which is of oscillatory nature and takes into account reflection of the Auger-electron with energy $\varepsilon$ by the $C_{60}$ shell. The details of calculation of $F_A(\varepsilon)$ can be found in Refs. [4, 5, 15]. In short, this function is expressed via the regular and irregular at $r = 0$ photoelectron wave functions.

Entirely, the following relation gives the Auger-decay amplitude $A_{i \to fk}^{C_{60}}$ for an atom inside the $C_{60}$ shell:

$$A_{i \to fk}^{C_{60}} \approx F_A(\varepsilon_k) A_{i \to fk}. \qquad (15)$$

Using this amplitude, one has for the Auger width, after neglecting the modification due to $C_{60}$ shell excitation, described by the second term in (12), the following relation:

$$\Gamma_{A,i \to fk}^{C_{60}} \approx |F_A(\varepsilon_k)|^2 \Gamma_{A,i \to fk}^A. \qquad (16)$$

The amplitude factor $|F_A(\varepsilon)|^2$ as a function of kinetic photoelectron energy for different encapsulated atoms was calculated in Refs. [4, 5, 15]. This is a rapidly variation function. The shape and positions of oscillations in $|F_A(\varepsilon)|^2$ are very sensitive to the concrete atom A encapsulated and to the magnitude of the fullerene shell radius $R$ [15].

**6. Results of calculating radiative and Auger widths**

The calculation results, according to formula (7), of a ratio between the width of the radiative vacancy decay in the A atom inside the fullerene shell and the width of the same decay in the free A atom $\eta_{RR}(\omega) = \Gamma_{\gamma,if}^{At} / \Gamma_{\gamma,if}^{C60}$ is given in Fig. 2. The frequency dependence $\eta_{RR}(\omega)$ is a curve with two pronounced maxima, for ~5 eV and ~25 eV.

The curve presented in Fig. 2 differs significantly from that obtained within the assumption that the fullerene shell can be considered as an ideally conducting sphere [10]. In the static limit, i.e. for zero radiation frequencies the function $\eta_{RR}(0) \approx 2.26$ while in the ideally conducting sphere approximation this function has to be equal to zero. When the frequency $\omega$ increases, the function $\eta_{RR}(\omega) \to 1$. The curve maxima reach values of about 6-7. According to Fig. 2, over the range of photon energies 0-50 eV the radiative widths of the vacancy decays in the endohedral and free atoms are significantly different and strongly depend on recombination photon energy.

The results obtained in this paper for $\eta_{RR}(\omega)$ differ in the area of small $\omega$ from that obtained for the same parameter in [8]. For $\omega$ close to $\eta_{RR}(\omega)$ maximum and higher our calculations here and in [8] are in agreement.

The reason for this is the following. In [8] results for $\eta_{RR}(\omega)$ were obtained using the dipole dynamic polarizability of $C_{60}$ calculated in Ref. [14]. There the cross-section of $C_{60}$ photoabsorption was approximating by a Lorenz symmetrical profile with the experimental value of half width of the Giant resonance. This profile was normalizing using the sum rule. The Lorenz shape is similar to that for ideally conducting sphere.

Here we used the real fulleren $C_{60}$ cross section that is asymmetric and is thus far from an ideal conducting sphere.

Calculated using formula (10), the ratio of the Auger-decay width of the vacancy in the endohedral atom to the radiative width of the free atom $\eta_{AR}(\omega) = \Gamma^{C60}_{A,if} / \Gamma^{A}_{\gamma,if}$ is given in Fig. 3. At the threshold of the fulleren shell $C_{60}$ ionization this ratio vanishes and $\Delta E_1 = 26.85$ eV corresponds to $2s$ - $2p$ transition in Ne;

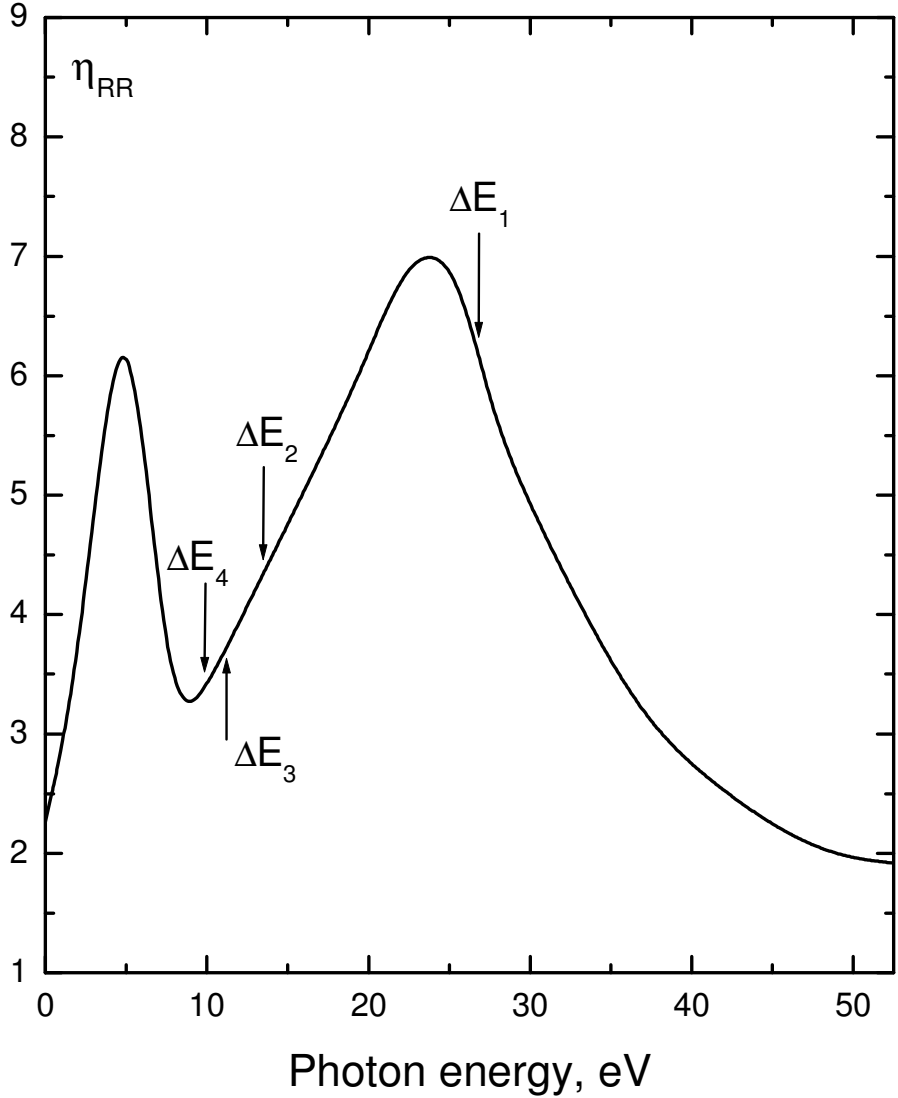

Fig. 2. The ratio $\eta_{RR}(\omega)$ of the radiative widths for the same transition in endohedral and free atom as a function of photon energy.

$\Delta E_2 = 13.50$ eV is $3s$ - $3p$ transition in Ar or $4s$ – $4p$ in Kr;

$\Delta E_3 = 11.20$ eV and $\Delta E_4 = 9.90$ eV are $5s$ - $5p_{3/2}$ and $5s$ - $5p_{1/2}$ transitions in Xe, respectively.

All transition energies are given in Ref. [16]

then rapidly reaches its maximal value $\eta_{AR}(5\,\text{eV}) \approx 6.3 \cdot 10^6$. With increase of transition frequency $\eta_{AR}(\omega)$ rapidly decreases. The function $\eta_{AR}(\omega)$ for transition energy within the range of $10 \leq \omega \leq 50\,\text{eV}$ is given in the insert of this figure. Thus, intershell interaction in the endohedral systems $A@C_{60}$ radically (by several orders of magnitude) increases the probability of the vacancy decay. This result is in agreement with the corresponding value for the $3s \to 3p$ transition in $Ne@C_{60}$ obtained in Ref. [17].

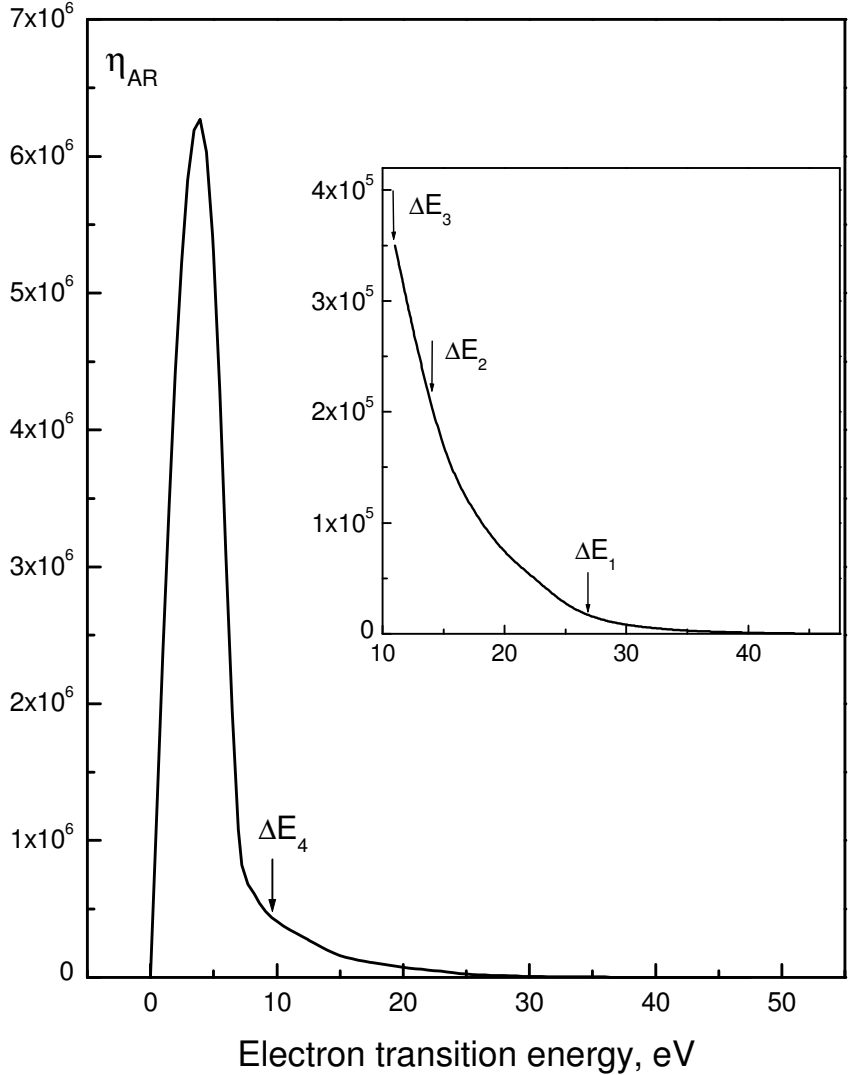

Fig. 3. The ratio of Auger and radiative widths $\eta_{AR}(\omega)$ as a function of transition energy. Arrows have the same means as in Fig. 2.

Calculated with formula (11), the fluorescence yield $J_R(\omega)$ as a function of electron transition energy is given in Fig. 4. This function goes to infinity at the threshold since the imaginary part of polarizability at this point is equal to zero and then rapidly increases with the raise in electron transition energy. In Fig. 4 we omitted the rapidly changing part of the curve near the threshold where the experimental data about the photodetachment cross-section cannot be considered to be sufficiently reliable. As in the previous case, the ratio of the widths is extremely sensitive to transition energy and the fluorescence yield changes by several orders within the range of transition energy $10 \leq \omega \leq 50\,\text{eV}$.

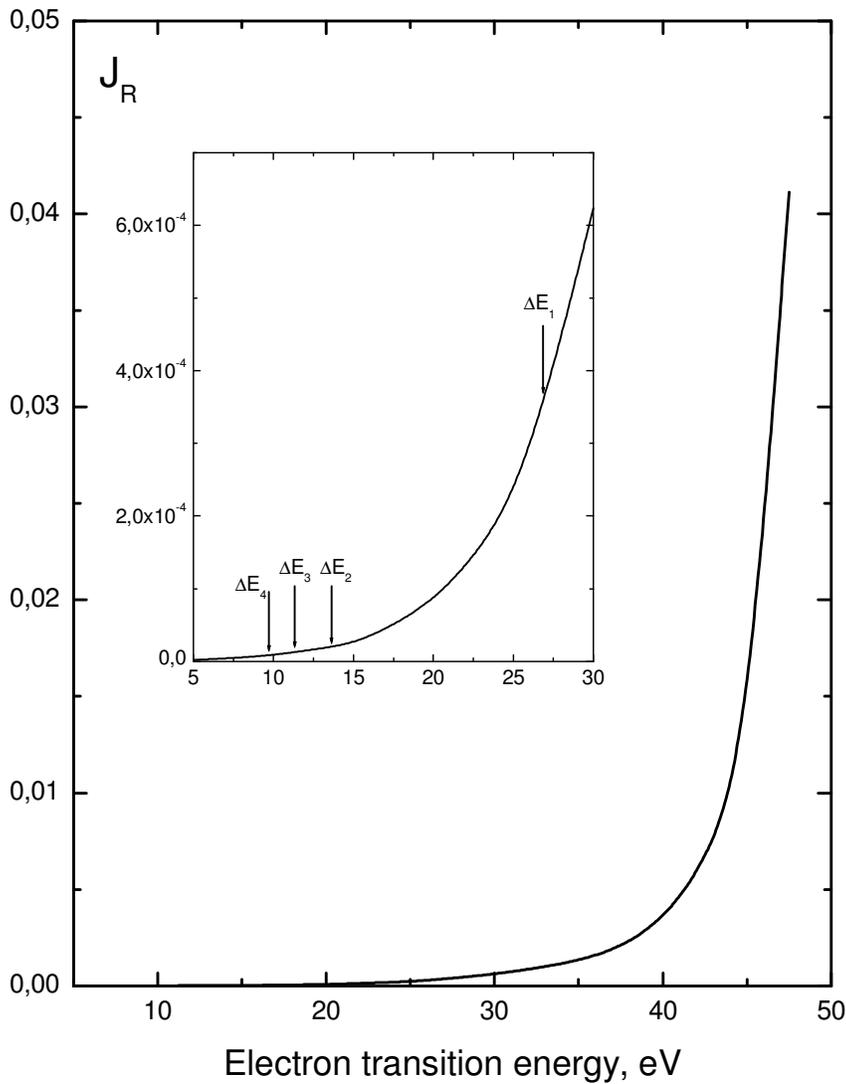

Fig. 4. The fluorescence yield $J_R(\omega)$ as a function of transition energy.

## 7. Conclusions

It was demonstrated in this paper that the dynamic reaction of the $C_{60}$ collectivized electrons radically changes the probabilities of the radiative and non-radiative vacancy decays in the endohedral atoms. Existence of a great number of the collectivized electrons in the fullerene shell and their participation in the processes of vacancy decays in the encapsulated atoms changes by several orders of magnitude the widths of these processes.

The small ratio of the atomic shell radius to that of $C_{60}$ shell provides the opportunity to obtain remarkably simple formulas for the probabilities of the electron transitions. This small parameter makes it possible to consider the electronic sub-systems of the inner atom and $C_{60}$ shell as weakly interacting and describe the effect of the spherical shell upon the electronic process in the A atom by several universal formulas.

This is the function $F_A(\varepsilon)$ [see Eq. (16)], built on the basis of the experimental data about energy of electron affinity to the empty fullerene shell and radius of this $C_{60}$-shell [8]. By this function are defined the shape of electronic spectra of photoionization of an endohedral atom and Auger-vacancy decay in this atom.

This is the function $\eta_{RR}(\omega)$ built on the basis of the experimental data about photoionization of the $C_{60}$ shell and making it possible to sequentially consider a dynamic reaction of the collectivized electrons both to the processes of endohedral atom photoionization and to those of radiative recombination.

And finally, this is the function $\eta_{AR}(\omega)$ describing the non-radiative collapse of the endohedral atom vacancy with simultaneous ionization of the fulleren shell. The remarkable moment here is that all considered here processes are parametrically correctly described with no consideration of any details of the electronic structure of the $C_{60}$ shell. Any attempt to take into account this structure is inevitably accompanied by radical simplifications, the correctness of which is very difficult to estimate on the pure theoretical grounds.

## 8. Acknowledgements

The authors are grateful for financial support to Bi-national Science Foundation, Grant 2002-064, Israeli Science Foundation, Grant 174/03. This work was also supported by Uzbekistan National Foundation, Grant Ф-2-1-12.